\documentclass[a4paper,11pt]{article}
\pdfoutput=1 

\usepackage{lineno}
\usepackage{jinstpub} 
\usepackage{xspace}
\usepackage{makecell}
\usepackage{multirow}

\usepackage{comment}

\usepackage{pbox}

\title{COHERENT Collaboration data release from the measurements of CsI[Na] response to nuclear recoils}

\newcommand{\Mephi}{a}
\newcommand{\Mephidesc}{\affiliation[\Mephi]{National Research Nuclear University MEPhI (Moscow Engineering Physics Institute), Moscow, 115409, Russian Federation}}
\newcommand{\Duke}{b}
\newcommand{\Dukedesc}{\affiliation[\Duke]{Department of Physics, Duke University, Durham, NC, 27708, USA}}
\newcommand{\TUNL}{c}
\newcommand{\TUNLdesc}{\affiliation[\TUNL]{Triangle Universities Nuclear Laboratory, Durham, NC, 27708, USA}}
\newcommand{\UTK}{d}
\newcommand{\UTKdesc}{\affiliation[\UTK]{Department of Physics and Astronomy, University of Tennessee, Knoxville, TN, 37996, USA}}
\newcommand{\ITEPnewa}{e}
\newcommand{\ITEPnewadesc}{\affiliation[\ITEPnewa]{National Research Center  ``Kurchatov Institute'' , Moscow, 123182, Russian Federation }}
\newcommand{\ORNL}{f}
\newcommand{\ORNLdesc}{\affiliation[\ORNL]{Oak Ridge National Laboratory, Oak Ridge, TN, 37831, USA}}
\newcommand{\USD}{g}
\newcommand{\USDdesc}{\affiliation[\USD]{Department of Physics, University of South Dakota, Vermillion, SD, 57069, USA}}
\newcommand{\NCSU}{h}
\newcommand{\NCSUdesc}{\affiliation[\NCSU]{Department of Physics, North Carolina State University, Raleigh, NC, 27695, USA}}
\newcommand{\Sandia}{i}
\newcommand{\Sandiadesc}{\affiliation[\Sandia]{Sandia National Laboratories, Livermore, CA, 94550, USA}}
\newcommand{\UW}{j}
\newcommand{\UWdesc}{\affiliation[\UW]{Center for Experimental Nuclear Physics and Astrophysics \& Department of Physics, University of Washington, Seattle, WA, 98195, USA}}
\newcommand{\LANL}{k}
\newcommand{\LANLdesc}{\affiliation[\LANL]{Los Alamos National Laboratory, Los Alamos, NM, 87545, USA}}
\newcommand{\Laurentian}{l}
\newcommand{\Laurentiandesc}{\affiliation[\Laurentian]{Department of Physics, Laurentian University, Sudbury, Ontario, P3E 2C6, Canada}}
\newcommand{\CMU}{m}
\newcommand{\CMUdesc}{\affiliation[\CMU]{Department of Physics, Carnegie Mellon University, Pittsburgh, PA, 15213, USA}}
\newcommand{\IU}{n}
\newcommand{\IUdesc}{\affiliation[\IU]{Department of Physics, Indiana University, Bloomington, IN, 47405, USA}}
\newcommand{\VT}{o}
\newcommand{\VTdesc}{\affiliation[\VT]{Center for Neutrino Physics, Virginia Tech, Blacksburg, VA, 24061, USA}}
\newcommand{\NCCU}{p}
\newcommand{\NCCUdesc}{\affiliation[\NCCU]{Department of Mathematics and Physics, North Carolina Central University, Durham, NC, 27707, USA}}
\newcommand{\UF}{q}
\newcommand{\UFdesc}{\affiliation[\UF]{Department of Physics, University of Florida, Gainesville, FL, 32611, USA}}
\newcommand{\Tufts}{r}
\newcommand{\Tuftsdesc}{\affiliation[\Tufts]{Department of Physics and Astronomy, Tufts University, Medford, MA, 02155, USA}}
\newcommand{\SNU}{s}
\newcommand{\SNUdesc}{\affiliation[\SNU]{Department of Physics and Astronomy, Seoul National University, Seoul, 08826, Korea}}
\author[\Mephi]{D.~Akimov,}\Mephidesc
\author[\Duke,\TUNL]{P.~An,}\Dukedesc\TUNLdesc
\author[\Duke,\TUNL]{C.~Awe,}
\author[\Duke,\TUNL]{P.S.~Barbeau,}
\author[\UTK]{B.~Becker,}\UTKdesc
\author[\Mephi,\ITEPnewa]{V.~Belov ,}\ITEPnewadesc
\author[\UTK]{I.~Bernardi,}
\author[\ORNL]{M.A.~Blackston,}\ORNLdesc
\author[\USD]{C.~Bock,}\USDdesc
\author[\Mephi]{A.~Bolozdynya,}
\author[\NCSU]{J.~Browning,}\NCSUdesc
\author[\Sandia]{B.~Cabrera-Palmer,}\Sandiadesc
\author[\USD,1]{D.~Chernyak,}\note{Now at: Department of Physics and Astronomy, University of Alabama, Tuscaloosa, AL, 35487, USA and Institute for Nuclear Research of NASU, Kyiv, 03028, Ukraine}
\author[\Duke]{E.~Conley,}
\author[\ORNL]{J.~Daughhetee,}
\author[\UW]{J.~Detwiler,}\UWdesc
\author[\USD]{K.~Ding,}
\author[\UW]{M.R.~Durand,}
\author[\UTK,\ORNL]{Y.~Efremenko,}
\author[\LANL]{S.R.~Elliott,}\LANLdesc
\author[\ORNL]{L.~Fabris,}
\author[\ORNL]{M.~Febbraro,}
\author[\Laurentian]{A.~Gallo Rosso,}\Laurentiandesc
\author[\ORNL,\UTK]{A.~Galindo-Uribarri,}
\author[\TUNL,\ORNL,\NCSU]{M.P.~Green,}
\author[\ORNL]{M.R.~Heath,}
\author[\Duke,\TUNL]{S.~Hedges,}
\author[\CMU]{D.~Hoang,}\CMUdesc
\author[\IU]{M.~Hughes,}\IUdesc
\author[\Duke,\TUNL]{T.~Johnson,}
\author[\Mephi]{A.~Khromov,}
\author[\Mephi,2]{A.~Konovalov,}\note{Now at: Lebedev Physical Institute of the Russian Academy of Sciences: 53 Leninskiy Prospekt, Moscow, 119991, Russian Federation}
\author[\Mephi]{E.~Kozlova,}
\author[\Mephi]{A.~Kumpan,}
\author[\Duke,\TUNL]{L.~Li,}
\author[\VT]{J.M.~Link,}\VTdesc
\author[\USD]{J.~Liu,}
\author[\NCSU]{K.~Mann,}
\author[\NCCU,\TUNL]{D.M.~Markoff,}\NCCUdesc
\author[\IU]{J.~Mastroberti,}
\author[\ORNL]{P.E.~Mueller,}
\author[\ORNL]{J.~Newby,}
\author[\CMU]{D.S.~Parno,}
\author[\ORNL]{S.I.~Penttila,}
\author[\Duke]{D.~Pershey,}
\author[\CMU,3]{R.~Rapp,}\note{Now at: Washington \& Jefferson College, Washington, PA, 15301, USA}
\author[\UF]{H.~Ray,}\UFdesc
\author[\Duke]{J.~Raybern,}
\author[\Mephi,\ITEPnewa]{O.~Razuvaeva,}
\author[\Sandia]{D.~Reyna,}
\author[\TUNL]{G.C.~Rich,}
\author[\NCCU,\TUNL]{J.~Ross,}
\author[\Mephi]{D.~Rudik,}
\author[\Duke,\TUNL]{J.~Runge,}
\author[\IU]{D.J.~Salvat,}
\author[\CMU]{A.M.~Salyapongse,}
\author[\USD]{J.~Sander,}
\author[\Duke]{K.~Scholberg,}
\author[\Mephi]{A.~Shakirov,}
\author[\Mephi,\ITEPnewa]{G.~Simakov,}
\author[\Duke,4]{G.~Sinev,}\note{Now at: South Dakota School of Mines and Technology, Rapid City, SD, 57701, USA}
\author[\IU]{W.M.~Snow,}
\author[\Mephi]{V.~Sosnovtsev,}
\author[\IU]{B.~Suh,}
\author[\IU]{R.~Tayloe,}
\author[\VT]{K.~Tellez-Giron-Flores,}
\author[\IU,5]{I.~Tolstukhin,}\note{Now at: Argonne National Laboratory, Argonne, IL, 60439, USA}
\author[\NCCU,\TUNL]{E.~Ujah,}
\author[\IU]{J.~Vanderwerp,}
\author[\ORNL]{R.L.~Varner,}
\author[\Laurentian]{C.J.~Virtue,}
\author[\IU]{G.~Visser,}
\author[\Tufts]{T.~Wongjirad,}\Tuftsdesc
\author[\USD]{Y.~Yang,}
\author[\CMU]{Y.-R.~Yen,}
\author[\SNU]{J.~Yoo,}\SNUdesc
\author[\ORNL]{C.-H.~Yu,}
\author[\IU,6]{J.~Zettlemoyer}\note{Now at: Fermi National Accelerator Laboratory, Batavia, IL, 60510, USA}
\emailAdd{amkonovalov@mephi.ru,daniel.pershey@duke.edu}

\emailAdd{}

\abstract{Description of the data release~\cite{Release_2023} from the measurements of the CsI[Na] response to low energy nuclear recoils by the COHERENT collaboration. The release corresponds to the results published in ref.~\cite{Akimov_2021}. We share the data in the form of raw ADC waveforms, provide benchmark values, and share plots to enhance the transparency and reproducibility of our results. This document describes the contents of the data release as well as guidance on the use of the data.}

\collaboration[c]{(The COHERENT collaboration)}

\begin{document}
\maketitle
\flushbottom


\section{Overview of the release} \label{sec:overview}

The data release~\cite{Release_2023} includes the CsI[Na] nuclear recoil quenching factor (QF) data acquired in a series of measurements performed by the COHERENT collaboration and described in ref.~\cite{Akimov_2021} and references therein. We duplicate the scripts from the release in the public repository~\cite{Gitlab_2023}, which can be updated if any issues are found. Please direct questions about the material provided within this release to \texttt{amkonovalov@mephi.ru} and \texttt{daniel.pershey@duke.edu}.

\section{Structure} \label{sec:structure}

The root directory of the data release is \texttt{csi\_qf\_data\_release}. It contains five subfolders. Four of them --- \texttt{COHERENT-1/2/3/4} --- correspond to COHERENT CsI[Na] QF measurements and the remaining \texttt{global\_qf\_data\_fit} is for a global QF data fit tool based on the data from ref.~\cite{Park_2002, Guo_2016, Collar_2019, Akimov_2021}. Each of the numbered \texttt{COHERENT} folders contains the data organized according to the steps in our data acquisition and analysis:

\begin{itemize}
  \item \texttt{n\_beam\_energy\_tof} --- time of flight data for characterizing the energy distribution of incident neutrons;
  \item \texttt{BD\_calibration} --- data for the calibration of the backing detectors (BD);
  \item \texttt{CsI\_calibration} --- data for the calibration of the CsI[Na] detector;
   \item \texttt{n\_beam\_data} --- neutron beam data for determining the CsI[Na] QF.
\end{itemize}

Example code macros to read the data files and plot the recorded waveforms are also stored in the \texttt{COHERENT} folders and have names similar to \texttt{COH?\_viewer.*} regular expression. \texttt{COHERENT} folders also contain the \texttt{*\_neutron\_beam\_energy.*} files with the energy distributions of incident neutrons evaluated from the time-of-flight data. The MCNPX-PoliMi based predictions of the nuclear recoil energy depositions in CsI[Na] and macros to read these predictions are stored in the \texttt{nuclear\_recoil\_prediction} subfolders inside \texttt{n\_beam\_data} folders.

\section{Guidance on analysis} \label{sec:guidance}

\subsection{Evaluation of incident neutron energy} \label{subsec:neutron_en}

The incident neutron energy is evaluated based on the time of flight data acquired with an EJ~-~309 liquid scintillator detector. The detector was placed at a known distance from the ion beam target -- either a deuterium gas cell or Li foil -- in which neutrons were produced. Beam related gamma rays and neutrons are registered by the detector. The capability of EJ-309 to distinguish between signals from neutrons and gamma rays provides an estimate of the delay between the arrival of the former and the latter, enabling the kinematic reconstruction of incident neutron energies. Both arrival times are defined relative to the periodic signal of the beam-pick-off monitor (BPM) associated with primary ion beam pulse. The exact expression for the velocity of neutrons can be evaluated in the following way:

\begin{equation}
\label{eq:tof1}
\begin{gathered}
  ct = v_{n}t+v_{n}\Delta t = d,\\
  v_{n}=\frac{cd}{c \Delta t+d},
\end{gathered}
\end{equation}
where $d$ is the distance travelled by neutrons, $t$ is the time needed for a gamma ray to travel $d$, $\Delta t$ is the time delay between arrival times of neutrons and gamma rays and $c$ is the speed of light. We suggest converting directly from the time of flight spectrum (histogram)
to a distribution of neutron velocities and then to energy while taking into account non-linear dependence of the energy on velocity (variable bin size of the final histogram). Results of such a straightforward approach for COHERENT-1/2 match with the more complex evaluation based on the MCNPX-PoliMi simulation. We recommend using constant fraction discrimination (CFD) thresholds for analysis of both the EJ-309 and BPM waveforms. It is useful to pick only one of the BPM pulses (e.g. the first following/preceding the EJ-309 signal) to determine the time delay if several pulses are presented in the recorded waveform. The leading uncertainties in neutron energies are associated with the time resolution of the measurements, the number of distances at which measurements were performed, and uncertainties in the neutron production site in the source as well as interaction site in the EJ-309 cell. Table~\ref{Table_Beam} presents a summary of the available time of flight measurements. The distances listed take into account the contributions from the source geometries and the EJ-309 cell. The uncertainty of the distance measurements is about 3.8~cm for COHERENT-1/2/3 and about 1~cm for COHERENT-4 for the absolute value, but is cancelled out if the difference of distances is considered. The difference in path length between gamma rays and neutrons is negligible. The velocity of neutrons can be calculated with the classical expression. Examples of time-of-flight spectra and evaluated neutron energy distributions can be found in Fig.~1 and Tab.~2 of ref.~\cite{Akimov_2021}. The description of the MCNPX-PoliMi simulation of the TOF data and the corresponding neutron beam energy spectra for COHERENT-1/2 can be found in Appendix D of ref.~\cite{Grayson_2017}. The definitions of pulse shape discrimination (PSD) parameters required to separate neutron- and gamma-induced signals in the EJ-309 detector and illustrations of PSD plots can be found in Fig. 3, 6, 9 of ref.~\cite{Akimov_2021}.

\begin{table}
\begin{center}
\parbox{0.85\linewidth}{\caption{\label{Table_Beam} Incident neutron TOF data}}
\begin{tabular}{lcccc}
\hline
Dataset    & Mean $E_{n}$, MeV & Stand-off distances, cm & Measured by & Time sample, ns \\ \hline
COHERENT-1/2 & 3.8 & 387.1, 518.2, 570.6 & TAC & $\sim0.11$ \\
\hline
COHERENT-3 & 4.4 & 356.2, 417.2, 449.4 & ADC & 4 \\
\hline
\multirow{2}{*}{COHERENT-4} & 0.94 & 138.2 & \multirow{2}{*}{ADC} & \multirow{2}{*}{4} \\
& 1.26 & 150.3, 194.7, 217.6 & & \\
\hline
\end{tabular}
\end{center}
\end{table}

The neutron energy can be verified by time delays between neutrons and gammas at different distances from the source. Such a cross-check is independent from the absolute measurement of the distance from the source to the EJ-309 and uses relative distance information for different positions of the detector:

\begin{equation}
  \label{eq:tof2}
  v_{n}=\frac{d_{2}-d_{1}}{t_{2}-t_{1}},
\end{equation}
where $d_{1}, d_{2}$ are distances between the source and EJ-309 and $t_{1}, t_{2}$ are maxima of neutron arrival time distributions at these distances. Such a method suffers from increased relative uncertainty because of the compound effects of two distance measurements.

\subsection{Calibration of backing detectors} \label{subsec:bd_calib}

The COHERENT-1/2/3 QF measurements utilized organic scintillator detectors to tag neutrons scattering off of the CsI[Na] crystal. The calibration of the electron recoil (ER) energy scale of these backing detectors is required to take into account the effect BD energy deposition selections have on nuclear recoil (NR) spectra. We calibrated all 12 EJ-309 BDs of the COHERENT-1 measurement with a neutron energy deposition having an endpoint of about 1.3 MeV$_{ee}$ \cite{Pino_2014}. For the COHERENT-2 measurement we suggest calibrating the EJ-299-33A BD with $^{22}$Na and $^{137}$Cs gamma rays of 511 keV and 662 keV with corresponding Compton edges of 341 keV and 477 keV. Plots presenting the results of such a calibration can be found in ref.~\cite{Scholz_2017} (Fig. 8.5). The calibration of the BD energy scale coincide in the original analysis~\cite{Scholz_2017} and reanalysis~\cite{Akimov_2021} within 3\%. The $^{252}$Cf calibration provides verification of the PSD parameter values for neutron-induced NRs and signals from gamma rays. The $^{137}$Cs source was also used to calibrate the EJ-309 liquid scintillator BD in the COHERENT-3 measurement.

\subsection{Calibration of CsI[Na]} \label{subsec:csi_calib}

\begin{table}
\begin{center}
\parbox{1.0\linewidth}{\caption{\label{Table_ADC} ADCs, integral units and benchmark values}}
\begin{tabular}{lcccc}
\hline
Dataset & COHERENT-1 & COHERENT-2 & COHERENT-3 & COHERENT-4 \\ \hline
ADC                 & CAEN V1730 & Acquiris U1071A & SIS3316 & SIS3316 \\
Resolution, bits    & 14  & 8 & 14 & 14 \\
Dynamic range, V    & 2.0 & 0.05 & 2.0 & 2.0 \\
Sampling rate, MS/s & 500 & 500 & 250 & 250 \\
SPE, ADC units & $1300\pm20$ & $49.0\pm1.2$ & $64.5\pm1.5$ & $86.0\pm1.7$\\
59.5 keV peak, $\times10^{3}$ ADC units & $1080\pm10$ & $48.1\pm0.5$ & $44.0\pm0.4$ & $57.5\pm0.4$\\
59.5 keV peak, nVs & $264\pm2$ & $18.8\pm0.2$ & $21.5\pm0.2$ & $28.1\pm0.2$\\
\end{tabular}
\end{center}
\end{table}

The calibration of the CsI[Na] ER energy scale was performed with the 59.5 keV gamma ray line of $^{241}$Am in all of the COHERENT QF measurements. The signal analysis approach description and illustrations of the energy deposition spectra can be found in Section 3 and Fig. 2 of ref.~\cite{Akimov_2021}. Table~\ref{Table_ADC} presents benchmark values of the 59.5 keV peak positions in ``ADC units'', which is a measure of integral equal to ``ADC counts $\times$ ADC time sample'' for each measurement. We also provide information needed for the conversion from ADC units to nVs and PMT photoelectrons~(PE). The uncertainties in Tab.~\ref{Table_ADC} represent rather the spread of the values obtained throughout the dataset, than particular statistic or systematic uncertainty. Note that the mean single photoelectron~(SPE) integrals presented in Tab.~\ref{Table_ADC} were evaluated with the help of a Gaussian fit. Such a fit does not provide consistent description of the mean SPE integral at different PMT bias voltages and can induce bias relative to the true SPE value (see discussions in Appendix B of ref.~\cite{Akimov_2021}). The absolute value of the SPE mean integral does not affect the evaluated QF values as it cancels out in the definition of the QF. The second order dependence may come from the effect of the absolute light yield on the resolution model and smearing of the predicted NR spectrum.

The 59.5 keV peak position for COHERENT-1 from Tab.~\ref{Table_ADC} differs from the one suggested by ref.~\cite{Grayson_2017}. The latter had an issue with onset finding in $^{241}$Am signals leading to ~3\% lower 59.5 keV peak position and distorted spectral shape. The issue is confirmed by the original author of ref.~\cite{Grayson_2017}. The 59.5 keV response after fixing the issue in the original analysis pipeline coincides with the one presented in Tab.~\ref{Table_ADC}. We also note that ref.~\cite{Grayson_2017} used the Polya model of the mean SPE integral. We use a Gaussian model here and in ref.~\cite{Akimov_2021} for consistency between data-sets and comparison with the results of ref.~\cite{Collar_2019}. The SPE spectra presented in Appendix B of ref.~\cite{Akimov_2021} suggest that the Polya model alone cannot describe the full SPE integral distribution.

The 59.5 keV peak characteristic integral coincides in the original analysis of COHERENT-2~\cite{Scholz_2017} and reanalysis presented in ref.~\cite{Akimov_2021} to within 1\%, although the original analysis used the units of "V$\times$2ns" (twice smaller than nVs from Tab.~\ref{Table_ADC}). The SPE mean estimates obtained with the Gaussian model in the ref.~\cite{Scholz_2017} are smaller by about 7\% than those in the re-analysis and provide the absolute light yield estimate of 17.7~PE/keV. We were unable to track down the reason for this discrepancy, but low energy signal integral estimates are different in these analyses in general as suggested by differing QF estimates.

\subsection{Evaluation of CsI[Na] QF} \label{subsec:csi_qf}

\begin{table}
\begin{center}
\parbox{0.85\linewidth}{\caption{\label{Table_sel} Requirements to select neutron-beam-related NR}}
\begin{tabular}{lcccc}
\hline
Selection & COHERENT-1 & COHERENT-2 & COHERENT-3 & COHERENT-4 \\ \hline
CsI pretrace cut & \checkmark & \checkmark & \checkmark & \checkmark \\
BD PSD and integral & \checkmark & \checkmark & \checkmark & n/a \\
BD-BPM timing    & \checkmark  & n/a & \checkmark & n/a \\
CsI-BD timing    & \checkmark & \checkmark & \checkmark & n/a \\
CsI-BPM timing   & $\times$ & n/a & $\times$ & \checkmark \\
CsI Cherenkov cut & $\times$ & $\times$ & \checkmark & \checkmark \\
\end{tabular}
\end{center}
\end{table}

The CsI[Na] QF evaluation relies on selection of the NR signals related to the neutron source pulses. Such selections utilize information from the backing detectors (in the tagged neutron measurements) and the beam-pick-off monitor. The general image of the selection requirements is presented in Tab.~\ref{Table_sel}, where ``\checkmark'' means that a requirement was used in the COHERENT analysis of a particular dataset, ``$\times$'' means that a requirement is available, but was not used, and ``n/a'' means that the requirement is not available due to absence of one of the signals needed. The illustrations of the distributions of restricted parameters can be found in corresponding sections of ref.~\cite{Akimov_2021}. The tagged neutron experiments rely on the selection of the neutron-induced signals of the BDs identified by the PSD parameter along with a BD integral cut. Additional accidental background suppression comes from the analysis of the time delay between BD and BPM signals (COHERENT-1 and COHERENT-3) or between BD and CsI[Na] signals. It is important to remember that the cuts relying on the onset of CsI[Na] signals (e.g. CsI-BD delay) may introduce inefficiency in selection of lowest energy signals due to the relatively slow scintillation decay times of the crystal. The COHERENT-4 measurement relies on the selection of neutron-induced signals of CsI[Na] by the time delay between CsI[Na] and BPM. In the COHERENT analysis additional effort was put into the suppression of Cherenkov-like signals in COHERENT-3 and COHERENT-4 data (described in corresponding sections of ref.~\cite{Akimov_2021}).

The empirical NR distributions can be fit to the prediction based on the MCNPX-PoliMi simulations. The prediction should account for the experimental resolution of the CsI[Na] setup with the leading contributions from the PMT photoelectron statistical fluctuations and fluctuations in the SPE integral. The afterglow contribution should also be taken into account if its contribution is non-negligible given the CsI[Na] pretrace cut (see Appendix A.3 and Tab.~8 in ref.~\cite{Akimov_2021}). The MCNPX-PoliMi predictions in \texttt{n\_beam\_data/nuclear\_recoil\_prediction} folders represent the prediction of the NR spectrum with none of the above mentioned resolution effects taken into account. The NR spectra were produced for the BD integral cuts described in ref.~\cite{Akimov_2021} for the COHERENT-1/2/3 data. Although the elastic scattering NR mean energy depends only weakly on the BD energy deposition, the contribution from the inelastic scattering of neutrons with gamma ray escape may affect NR mean energy, especially for the largest scattering angles of COHERENT-1. We therefore recommend using the BD integral cuts utilised in our analysis \cite{Akimov_2021}.

\subsection{Global CsI[Na] QF data fit} \label{subsec:csi_qf}

Together with the COHERENT CsI[Na] QF data we release the macro \texttt{qf\_csina\_fit\_PCA.C} to perform the global QF data fit. The input QF data are harmonized with ref.~\cite{Akimov_2021,Akimov_cevns} and allow the user to perform fits under different assumptions, taking into account sub-selections of the existing data-sets. Such a tool will be useful if any new measurements or clarifications on systematic uncertainties of the existing data appear.

\section{Verification and discussion of results}

In order to make the results of ref.~\cite{Akimov_2021} falsifiable we provide the benchmark lists of triggers from the CsI[Na] NR signal subselections (see \texttt{COHERENT-?/results} folders). These lists allow the unique identification of a trigger within a dataset and contain reconstructed event parameters used to evaluate CsI[Na] QF. We provide the images of waveforms for events from these lists to visualize our results or scripts to produce such images. The demonstrator event reconstruction scripts reproducing parameters from the lists are also released. Thus a way to verify both the reconstruction procedure and event selections of ours is provided. Interested parties may contact COHERENT by the e-mail addresses listed in this description for discussions. Communication will be supported for two years starting from the date of the data release.

\section{Summary} \label{sec:summary}

This paper describes the structure of the CsI[Na] QF data release by COHERENT~\cite{Release_2023} and presents the scripts (also in the repository~\cite{Gitlab_2023}) required to start working with the data. The global QF data fit scripts allow for the reproduction of the results used to estimate the coherent elastic neutrino-nucleus scattering cross-section in ref.~\cite{Akimov_cevns} and permit including new measurements or additional information on the systematic uncertainties in the fit. We hope that this release will contribute to enhancing the reproducibility of the CsI[Na] QF measurements. We encourage the scientific community working on the QF measurements to share the data, as existing measurements for the same materials indicate significant scatter, making it hard to reach consensus on the QF values and potential systematic uncertainty.

\section{Acknowledgements} \label{sec:aknow}

The COHERENT collaboration acknowledges the resources generously provided by the Spallation Neutron Source, a DOE Office of Science User Facility operated by the Oak Ridge National Laboratory. The Triangle Universities Nuclear Laboratory is supported by the U.S. Department of Energy under grant DE-FG02-97ER41033. This work was supported by the Ministry of Science and Higher Education of the Russian Federation, Project "New Phenomena in Particle Physics and the Early Universe" FSWU-2023-0073); the Russian Foundation for Basic Research (proj. \# 17-02-01077 A); the DOE HEP grant DE-SC0020518. Support for DOI 10.13139/OLCF/1969085 dataset is provided by the U.S. Department of Energy, project HEP106 under Contract DE-AC05-00OR22725. Project HEP106 used resources of the Oak Ridge Leadership Computing Facility at Oak Ridge National Laboratory, which is supported by the Office of Science of the U.S. Department of Energy under Contract No. DE-AC05-00OR22725. We are thankful to J. Collar and B. Scholz for their contribution to the COHERENT-2 measurement.

\appendix

\section{List of scripts} \label{sec:script_list}

In this section we present the list of scripts published with the release~\cite{Release_2023} (and in the repository~\cite{Gitlab_2023}) allowing to start the analysis of COHERENT CsI[Na] QF data and verify the results. The guidance on usage and compilation~(when needed) of the scripts is provided in the code of the scripts.

\subsection*{COHERENT-1}

\noindent\texttt{COHERENT-1/COH1\_viewer.cc} --- a script to view raw ADC waveforms of individual recorded events (triggers) of COHERENT-1;

\noindent\texttt{COHERENT-1/n\_beam\_energy\_tof/coh1\_and\_coh2\_tof\_reader.C} --- a script to read the neutron time-of-flight data recorded to evaluate the neutron beam energy in COHERENT-1 and COHERENT-2;

\noindent\texttt{COHERENT-1/n\_beam\_energy\_tof/coh1\_and\_coh2\_tof\_calc.C} --- a script to estimate the energy of beam neutrons based on the difference between time-of-flight values of gamma rays and neutrons;

\noindent\texttt{COHERENT-1/n\_beam\_data/nuclear\_recoil\_prediction/view\_nr\_spectrum.C} --- a script to view the simulated predictions of nuclear recoil energy depositions for COHERENT-1;

\noindent\texttt{COHERENT-1/results/verify\_parameters.cc} --- a script reproducing the approach used to evaluate the key parameters of signal events (triggers) of COHERENT-1;

\noindent\texttt{COHERENT-1/results/alt\_verify/alt\_verify\_parameters.cc} --- an analog of the verification script utilizing a more straightforward approach to pulse finding and signal integration, the CsI[Na] integrals calculated with its help are lower by 1--2\% in average compared with the default approach. 

\subsection*{COHERENT-2}

\noindent\texttt{COHERENT-2/parsed/COH2\_viewer.cc} --- a script to view raw ADC waveforms of individual recorded events (triggers) of COHERENT-2;

\noindent\texttt{COHERENT-2/data\_to\_root/COH2\_data\_to\_root.cc} --- a script to parse the raw binary data of COHERENT-2 and produce the ROOT tree with raw waveforms;

\noindent\texttt{COHERENT-2/results/show\_selected\_wfs.cc} --- a script creating \texttt{png} images with the waveforms of events from the signal subselections of COHERENT-2;

\noindent\texttt{COHERENT-2/results/verify\_parameters.cc} --- a script reproducing the approach used to evaluate the key parameters of signal events (triggers) of COHERENT-2;

\noindent\texttt{COHERENT-2/parsed/baseline\_pathology/base\_pathology.cc} --- a script illustrating the fluctuations of the DC baseline of ADC and the selections used to reject the triggers with significant fluctuations from the analysis of COHERENT-2.

\subsection*{COHERENT-3}

\noindent\texttt{COHERENT-3/COH3\_viewer.C} --- a script to view raw ADC waveforms of individual recorded events (triggers) of COHERENT-3;

\noindent\texttt{COHERENT-3/results/verify\_parameters.cc} --- a script reproducing the approach used to evaluate the key parameters of signal events (triggers) of COHERENT-3;

\subsection*{COHERENT-4}

\noindent\texttt{COHERENT-4/COH4\_viewer.C} --- a script to view raw ADC waveforms of individual recorded events (triggers) of COHERENT-4;

\noindent\texttt{COHERENT-4/COH4\_neutron\_beam\_energy.C} --- a script to show the functional form of the neutron energy distributions in COHERENT-4;

\noindent\texttt{COHERENT-4/n\_beam\_data/0.94MeV/nuclear\_recoil\_prediction/read\_nr\_tree.cc} --- a script to view the simulated predictions of nuclear recoil energy depositions for the 0.94 MeV run of COHERENT-4;

\noindent\texttt{COHERENT-4/n\_beam\_data/1.26MeV/nuclear\_recoil\_prediction/read\_nr\_tree.cc} --- a script to view the simulated predictions of nuclear recoil energy depositions for the 1.26 MeV run of COHERENT-4;

\noindent\texttt{COHERENT-4/results/verify\_parameters.cc} --- a script reproducing the approach used to evaluate the key parameters of signal events (triggers) of COHERENT-4;

\subsection*{Global QF data fit}

\noindent\texttt{global\_qf\_data\_fit/qf\_csina\_fit\_PCA.C} --- a script to reproduce the global qf data fit evaluated in ref.~\cite{Akimov_2021} and used in ref.~\cite{Akimov_cevns}, contains QF and visible energy values as well as nuclear recoil energy with appropriate uncertainties.

\section{DC baseline voltage fluctuations of ADC in COHERENT-2} \label{sec:coh2_base}

The inspection of COHERENT-2 raw waveforms recorded by the Acquiris U1071A ADC suggests that about 20\% of the triggers suffer from significant fluctuations in the DC voltage baseline of the ADC. Such fluctuations may distort the integrals of the low energy signals of interest and affect evaluated QF values. In this section we describe the way we find and reject the triggers with DC voltage fluctuations.
For each of the ADC channels we define the baseline value estimates based on certain waveform intervals. The recorded amplitude values for such an interval are used to fill a histogram which is then fit to a Gaussian distribution. This fit is performed in the range of $\pm$3 ADC units from the sample with the most probable amplitude value. The fit result for a Gaussian mean is used as a local baseline estimate. Below we list the symbols for these estimates and their combinations:

\noindent$V^{CsI}_{Beg}$ --- the estimate based on the first microsecond of the CsI[Na] waveform, also used as a ``default'' baseline value for the CsI channel analysis;

\noindent$V^{CsI}_{End}$ --- the estimate based on the last microsecond of the CsI[Na] waveform;

\noindent$\Delta V^{CsI}$ --- the difference between $V^{CsI}_{Beg}$ and $V^{CsI}_{End}$;

\noindent$V^{EJ}_{Beg}$ --- the estimate based on the first microsecond of the EJ plastic scintillator waveform;

\noindent$V^{EJ}_{Def}$ --- the estimate based on 0.9 to 1.9 $\mu s$ pre-trigger region of the EJ plastic scintillator waveform, a ``default'' baseline value for the EJ channel analysis;

\noindent$V^{EJ}_{End}$ --- the estimate based on the last microsecond of the EJ plastic scintillator waveform;

\noindent$\Delta V^{EJ}$ --- the difference between $V^{EJ}_{Beg}$ and $V^{EJ}_{End}$.

\begin{figure}[htbp]
\centering
\includegraphics[width=1.0\textwidth]{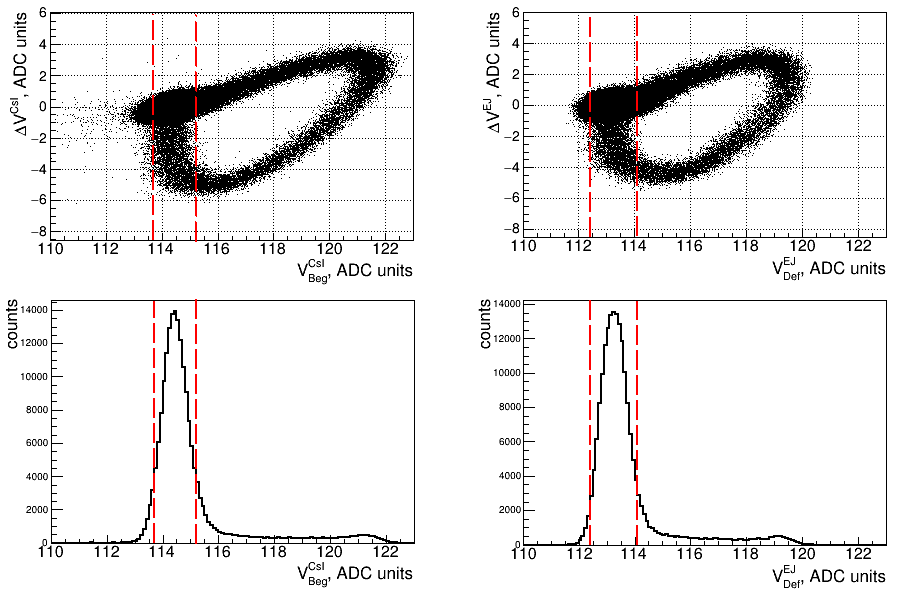}
\caption{\label{fig:BL_delta} Scatter plots illustrating fluctuations of the DC baseline voltage in the CsI[Na] and EJ channels (subselection of 45 degrees scattering angle data). Red lines show the cuts on the ``default'' baseline voltage values allowing to reject the majority of the distorted events.}
\end{figure}
\begin{figure}[htbp]
\centering
\includegraphics[width=1.0\textwidth]{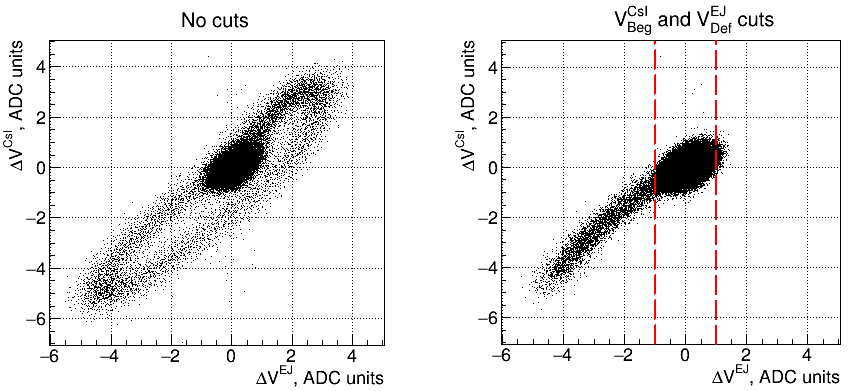}
\caption{\label{fig:BL_corr} Scatter plots illustrating correlation between $\Delta V^{CsI}$ and $\Delta V^{EJ}$ (subselection of 45 degrees scattering angle data). Left: no cuts; right: cuts on ``default'' baseline values of CsI[Na] and EJ channels. Red lines show an additional cut on the absolute value of $\Delta V^{EJ}$ allowing to reject remaining triggers with pathological fluctuations of the DC baseline.}
\end{figure}

We use a combination of these values calculated for each trigger with no ADC range overflows in either channel to characterize the baseline fluctuations. The correlation between the ``default'' baseline values and difference in the local baseline value estimates is presented in Figure \ref{fig:BL_delta}. Both channels demonstrate the image which can be interpreted as distortion and recovery of the baseline voltage. The maxima of the baseline value histograms correspond to absence of the $\Delta V$ with the exception of an excursion to the negative values. Another feature of the fluctuations is that $\Delta V^{CsI}$ is highly correlated with $\Delta V^{EJ}$, see Figure \ref{fig:BL_corr}. In order to reject the triggers with pathological baseline voltage fluctuations we, firstly, require $V^{CsI}_{Beg}$ and $V^{EJ}_{Def}$ to be contained within ~0.8 ADC units from the most probable value. Note that both these values are calculated based on the pre-trigger region and thus do not affect the expected nuclear recoil signals. Secondly, we restrict the absolute value of $\Delta V^{EJ}$ to be within 1 ADC unit. This latter cut allows us to reject the remaining pathological waveforms based on the correlation between $\Delta V^{CsI}$ and $\Delta V^{EJ}$. The tests show that $\Delta V^{EJ}$ doesn't depend on the size of recorded EJ signals up to events heavily affected by the ADC overflow. The EJ scintillation decay time is quite short and the signal does not reach the last microsecond of a recorded waveform, thus the cut doesn't introduce bias to the analysis of nuclear recoil signals from the beam neutrons. The remaining triggers demonstrate $\Delta V^{CsI}$ with a mean of about 0.02 ADC unit and RMS of 0.4 ADC units in the absence of signal. The average waveform accumulated from the triggers with no pulses in the CsI[Na] channel confirms absence of significant fluctuations in the baseline voltage value within a waveform. We thus conclude that remaining triggers are not affected by the pathological fluctuations of baseline voltage values and are suitable for the QF analysis. A script illustrating our selections can be found at \texttt{COHERENT-2/parsed/baseline\_pathology/base\_pathology.cc} within the release.


\begin{thebibliography}{40}

\bibitem{Release_2023}
COHERENT collaboration, \emph{COHERENT Collaboration data release from the measurements of CsI[Na] response to nuclear recoils}, {\bf 10.13139/OLCF/1969085} (2023), https://doi.ccs.ornl.gov/ui/doi/426

\bibitem{Akimov_2021}
COHERENT collaboration, \emph{Measurement of scintillation response of CsI[Na] to low-energy nuclear recoils by COHERENT},  \emph{Journal of Instrumentation} {\bf 17} (2022) P10034, arXiv:2111.02477

\bibitem{Gitlab_2023}
COHERENT collaboration, https://code.ornl.gov/COHERENT/qf\_data\_release

\bibitem{Park_2002}
H. Park et al., \emph{Neutron beam test of CsI crystal for dark matter search}, \emph{NIM A} {\bf491} (2002) 460

\bibitem{Guo_2016}
C. Guo et al., \emph{Neutron beam tests of CsI(Na) and CaF$_2$(Eu) crystals for dark matter direct search}, \emph{NIM A} {\bf818} (2016) 38

\bibitem{Collar_2019}
J.I. Collar, A.R.L. Kavner, and C.M. Lewis, \emph{Response of CsI[Na] to Nuclear Recoils: Impact on Coherent Elastic Neutrino-Nucleus Scattering (CE$\nu$NS)}, \emph{Physical Review D} {\bf 100} (2019) 033003, arXiv:1907.04828

\bibitem{Grayson_2017}
G.C. Rich, \emph{Measurement of Low-Energy Nuclear-Recoil Quenching Factors in CsI[Na] and Statistical Analysis of the First Observation of Coherent, Elastic Neutrino-Nucleus Scattering}, \emph{The University of North Carolina at Chapel Hill, PhD thesis} (2017)

\bibitem{Pino_2014}
F. Pino et al., \emph{The light output and the detection efficiency of the liquid scintillator EJ-309}, \emph{App. Radiation and Isotopes} {\bf89} (2014) 79

\bibitem{Scholz_2017}
B. Scholz, \emph{First Observation of Coherent Elastic Neutrino-Nucleus Scattering}, \emph{Springer International Publishing,  Springer Theses series} (2018)

\bibitem{Akimov_cevns}
COHERENT collaboration, \emph{Measurement of the Coherent Elastic Neutrino-Nucleus Scattering Cross Section on CsI by COHERENT},  \emph{Physical Review Letters } {\bf 129} (2022) 081801, arXiv:2110.07730

\end{thebibliography}
\end{document}